\pdfoutput=1
\documentclass[runningheads]{llncs}

\usepackage[T1]{fontenc}
\usepackage{lmodern}
\usepackage{graphicx}
\usepackage{hyperref}
\usepackage{amsmath,amssymb,mathtools}
\usepackage{booktabs}
\usepackage{xcolor}
\usepackage{tikz}
\usepackage{url}
\usepackage[activate={true,nocompatibility},final,tracking=true,
            kerning=true,spacing=true,factor=1100,stretch=20,shrink=20]{microtype}
\microtypecontext{spacing=nonfrench}
\usepackage[font=small,skip=4pt,labelfont=bf]{caption}

\usetikzlibrary{positioning,shapes.geometric,shapes.symbols,fit,arrows.meta,decorations.pathreplacing,calc,backgrounds}

\hypersetup{
  colorlinks=true,
  linkcolor=black,
  citecolor=black,
  urlcolor=blue,
  pdfauthor={Oliver Aleksander Larsen, Mahyar Tourchi Moghaddam},
  pdftitle={NostrAgent: A Decentralized Identity and Delegation Architecture for Sovereign Agentic Systems},
}

\usepackage{eso-pic}

\newcommand\copyrighttext{%
\footnotesize
This manuscript has been accepted as a Full Paper for presentation at the
AIPAA special session of the 2nd International Conference on Agentic and
Generative Techniques in Intelligent Computational Systems (AGENTICS) 2026,
held in Angers, France, 28 -- 30 October 2026, and for publication in
Springer Communications in Computer and Information Science (CCIS) proceedings.
This is the author's accepted manuscript version.
The final authenticated publication will be available via Springer.
}
\newcommand{\copyrightnotice}{%
\AddToShipoutPictureFG*{%
\AtPageLowerLeft{%
\raisebox{1cm}{%
\makebox[\paperwidth][c]{%
\fbox{\parbox[b]{0.9\textwidth}{\copyrighttext}}%
}}}}}

\begin{document}

\copyrightnotice

\title{NostrAgent: A Decentralized Identity and Delegation Architecture for Sovereign Agentic Systems}
\titlerunning{NostrAgent}

\author{Oliver Aleksander Larsen\inst{1} \and
Mahyar Tourchi Moghaddam\inst{1}}
\authorrunning{O. A. Larsen and M. T. Moghaddam}
\institute{University of Southern Denmark, Odense, Denmark\\
\email{\{olar,mtmo\}@mmmi.sdu.dk}}

\maketitle

\begin{abstract}
Autonomous AI agents increasingly act across organizational boundaries on behalf of human operators: they invoke third-party services, delegate subtasks to other agents, and pay for metered resources. Deploying such agents safely requires five capabilities that today live in separate systems: persistent identity, scoped delegation, peer trust, discovery, and payment. Existing approaches root these capabilities in centralized authorities or cover only subsets, so authority, trust, and payment fracture exactly where autonomy needs continuity: when a key rotates or a delegation must be revoked.
We present NostrAgent, a decentralized architecture that unifies all five capabilities over Nostr relays using three custom event kinds: Kind~38100 identity declarations authenticated by BIP340 Schnorr signatures with pre-rotation commitments, Kind~38101 scoped delegation chains whose every hop verifiably narrows the granted capabilities, and Kind~38102 peer attestations forming a Sybil-deterrent trust graph, with Lightning HTTP 402 (L402) binding payment to agent identity.
Identity remains operator-sovereign: the human operator controls it without any registration authority, relays are substitutable transport rather than a trust root, and every authorization decision is replayable offline from signed events, supporting interpretability and audit.
We evaluate a Python prototype with a mixed-method design: Architecture Tradeoff Analysis Method (ATAM) quality analysis with a two-round mini-Delphi expert panel, STRIDE threat modeling (Spoofing, Tampering, Repudiation, Information disclosure, Denial of service, Elevation of privilege) across three trust boundaries, eleven benchmarks with non-parametric statistics, and 19 failure modes.
Measured results show sub-millisecond offline verification, linear delegation-chain scaling, and end-to-end Lightning-settled L402 at 157~ms median on regtest. 17 of 19 failure modes pass empirical tests, one is bounded analytically, and one is disclosed as an architectural limitation. NostrAgent demonstrates an auditable prototype substrate for trustworthy agentic systems without centralized trust roots.

\end{abstract}

\section{Introduction}\label{sec:introduction}

Consider a logistics operator O that deploys an autonomous forecasting agent A. A discovers third-party weather-data services, spawns a retrieval sub-agent B and hands it a narrow slice of its own authority (read-only weather queries, expiring in 24 hours), and pays a data service S per query over Lightning. Every step crosses an organizational boundary, carries delegated authority, and moves money. When O suspects that A's key has been compromised, O must rotate it without losing B's delegation, A's accumulated peer reputation, or access to services A has already paid. The scenario runs through the paper: Section~\ref{sec:architecture} walks each mechanism through it, and Section~\ref{sec:evaluation} measures it.

Interactions of this kind demand five capabilities. \emph{Identity}: A and B must be persistently and verifiably identifiable to parties that have never met their operator. \emph{Delegation}: B must prove it acts within a scope A granted, and S must check that proof without calling anyone. \emph{Trust}: a newcomer choosing between competing weather agents needs evidence beyond self-description. \emph{Discovery}: A must find S without a central catalog that can exclude either of them. \emph{Payment}: cross-organizational agent services are metered in practice, as emerging agent-commerce rails assume~\cite{reppel2025x402,erc8183_2026}, and here payment is also a security mechanism: the per-identity cost floor makes peer trust expensive to fabricate (Section~\ref{sec:eval-rq2}). Identity is \emph{operator-sovereign} when the human operator controls it, its rotation, and its delegations without any external registration authority. Decentralized architecture is the means. Operator sovereignty is the property it guarantees.

Today these five capabilities live in separate systems: an OAuth authorization server, a DID registry with its credential specifications, a reputation service, a service catalog, and a billing rail. The seams between them are where autonomy breaks: each layer keeps its own key material, so one rotation traverses several unrelated lifecycles or silently orphans state; revocation does not cascade across layers; each layer brings a trust root to reconcile; and authorization needs online round-trips exactly where agents need local, offline decisions. Unifying the five capabilities on one substrate, one key type, one event log, and one revocation primitive removes these seams: authorization becomes a local check over signed events, and one event revokes an entire delegation subtree (Section~\ref{sec:evaluation}).

Recent analyses of agentic frameworks report a maturing but fragmented landscape with limited memory models, basic reasoning, and uneven ecosystem support, and find security largely underdeveloped~\cite{vaidhyanathan2026agentic}. Architecture evaluation methods for foundation-model agents likewise treat security as a first-class quality attribute~\cite{lu2026agentarceval}. We name the identity-layer deficits precisely:
\begin{itemize}
  \item \textbf{G1 (Key continuity):} rotate an agent key without losing identity, delegations, or accumulated trust.
  \item \textbf{G2 (Attenuated delegation):} pass authority onward such that every hop verifiably narrows scope, with cascading revocation.
  \item \textbf{G3 (Peer trust attestation):} establish capability-specific trust without a central reputation authority, at a cost that deters Sybil fabrication.
  \item \textbf{G4 (Registry-free discovery):} find agents and their endpoints without a registration authority.
  \item \textbf{G5 (Identity-bound payment):} pay for service access with credentials that are useless to a thief.
\end{itemize}
Existing systems cover subsets: OAuth~2.1~\cite{ietf2026oauth21}, AAP~\cite{ietf2026aap}, and SAGA~\cite{syros2026saga} root identity in a central provider (G4, and G1 becomes the provider's); W3C DIDs~\cite{w3c2022did} compose G2 and G3 from separate specifications; KERI/ACDC~\cite{smith2019keri,toip2023acdc} is strongest on G1 and G2 but offers no relay discovery or payment binding; AIP's self-certifying mode~\cite{prakash2026aip} lacks rotation, attestation, and payment; UCAN~\cite{ucan_spec_2025} attenuates capabilities without G1, G3, or G5; emerging IETF WIMSE (Workload Identity in Multi System Environments)~\cite{ietf_wimse_agent2026}, NIST~\cite{nist_nccoe_agents2026}, and OWASP~\cite{owasp_agentic_top10_2026} efforts address slices in parallel. Section~\ref{sec:related-work} substantiates this per system (Table~\ref{tab:comparison}). No surveyed system provides G1 through G5 on one substrate, and Section~\ref{sec:eval-rq2} evaluates against exactly these gaps.

We investigate three research questions, answered in Section~\ref{sec:evaluation}. \textbf{RQ1} asks how NostrAgent's design choices trade among sovereignty, operability, and trust. \textbf{RQ2} asks whether unifying identity, trust, and payment yields architectural advantages over composing separate systems. \textbf{RQ3} asks what failure modes arise in an operator-sovereign, relay-based architecture.

This paper makes the following contributions:
\begin{enumerate}
  \item[\textbf{C1.}] A unified operator-sovereign identity architecture with three custom Nostr event kinds (Kind~38100 identity, Kind~38101 delegation, Kind~38102 peer attestation), relay-based discovery via NIP-01 filters, pre-rotation key commitment, and L402 Lightning payment integration (Section~\ref{sec:architecture}; evaluated against G1-G5 in Section~\ref{sec:eval-rq2}).
  \item[\textbf{C2.}] Verifier-checked attenuated delegation chains in which each hop verifiably narrows scope, time bounds, and depth, with cascade revocation through parameterized replaceable events (Section~\ref{sec:arch-delegation}).
  \item[\textbf{C3.}] A mixed-method evaluation combining reduced-scope ATAM~\cite{kazman2000atam,sahlabadi2022lightweight} with a two-round mini-Delphi~\cite{dalkey1963delphi} expert panel, benchmarks across 37 metrics, STRIDE threat modeling~\cite{shostack2014threat}, and 19 failure modes (Section~\ref{sec:evaluation}).
  \item[\textbf{C4.}] An open-source Python prototype covering the full architecture on regtest Lightning, released as a reproducibility artifact (Section~\ref{sec:arch-prototype}; Data Availability).
\end{enumerate}

\section{Background}\label{sec:background}

\subsection{Agents, Operators, and the Protocol Ecosystem}\label{sec:bg-ecosystem}

An \emph{agent} is an autonomous software process, today typically driven by a large language model, that plans and acts on behalf of a human or organizational \emph{operator}. \emph{Delegation} passes a subset of an agent's authority to another agent. \emph{Attenuation} is the requirement that each hand-off narrows that authority and never widens it. The running example instantiates these roles: operator O, delegating agent A, delegatee B, service S.

Two protocols dominate how agents interact. The Model Context Protocol (MCP)~\cite{anthropic2024mcp} standardizes how an agent invokes tools and resources; Agent2Agent (A2A)~\cite{a2a2025spec} standardizes messaging between agents. Neither specifies who an agent \emph{is}, what authority it carries, or how it pays. NostrAgent is compositional with both rather than competing: Kind~38100 identity declarations carry MCP and A2A endpoint URLs (Section~\ref{sec:arch-identity}), so a NostrAgent identity wraps the protocols agents already speak. Table~\ref{tab:terminology} collects recurring terms.

\begin{table}[t]
\centering
\caption{Recurring terms used before or beyond their defining subsection.}
\label{tab:terminology}
\footnotesize
\begin{tabular}{@{}lp{8.3cm}@{}}
\toprule
Term & Meaning \\
\midrule
\texttt{d}-tag & Operator-chosen stable identifier keying a replaceable event \\
Preimage & Secret whose SHA256 hash locks a payment; proof of settlement \\
Satoshi (sat) & Smallest Bitcoin unit; Lightning payments are denominated in sats \\
Regtest & Local, deterministic Bitcoin/Lightning test network \\
Pre-rotation & Commitment to the hash of the next key before that key is ever used \\
Noisy-OR & Combination rule treating each trust path as independent evidence \\
ATAM & Architecture Tradeoff Analysis Method; scenario-based quality evaluation \\
STRIDE & Threat taxonomy over six categories (the rows of Table~\ref{tab:stride}) \\
B, FM, TP, SP & Benchmark, failure mode, trade-off point, sensitivity point (Section~\ref{sec:evaluation}) \\
\bottomrule
\end{tabular}
\end{table}

\subsection{BIP340 Schnorr Signatures}\label{sec:bg-bip340}

BIP340~\cite{wuille2020bip340} defines Schnorr signatures over the secp256k1 elliptic curve.
Public keys use a compact 32-byte x-only representation, and signatures are 64 bytes ($R \| s$). Verification is deterministic.
Because Bitcoin, Lightning, and Nostr all build on secp256k1, NostrAgent can bind a BIP340 identity key to a Lightning-settled macaroon without introducing a new trust root.

\subsection{Nostr Relay Model}\label{sec:bg-nostr}

Nostr is a relay-based protocol where clients exchange JSON events over WebSocket connections to relay servers~\cite{nostr2020nip01}. Each event carries a BIP340 signature from its author, so relays can store and forward events but cannot forge them. Event kinds 30\,000 through 39\,999 are \emph{parameterized replaceable events}, keyed by the triple (pubkey, kind, \texttt{d}-tag)~\cite{nostr2020nip01}. A newer event with the same key supersedes the prior one. Relays do not synchronize with each other, so agents must publish to every desired relay independently: relays are substitutable transport, not a trust root. We adopt Nostr because it co-locates BIP340 identity with that substitutable transport.

\subsection{L402: Lightning-Gated Authentication}\label{sec:bg-l402}

Macaroons~\cite{birgisson2014macaroons} are bearer credentials whose scope is constrained by HMAC-chained caveats. The Lightning Network~\cite{poon2016lightning} enables trustless micropayments via hash time-locked contracts. L402~\cite{osuntokun2020l402} combines these primitives by pairing a macaroon with a Lightning invoice. The client pays, obtains the preimage, and presents both as a unified credential. Embedding the agent's BIP340 public key in the macaroon identifier binds the credential to the requesting agent, an integration detailed in Section~\ref{sec:arch-l402}.

\section{Related Work}\label{sec:related-work}

Table~\ref{tab:comparison} compares ten representative systems across seven architectural axes chosen to operationalize the gaps of Section~\ref{sec:introduction}, so that the comparison and the evaluation measure the same things: A1 decentralized identity (the sovereignty predicates P1 through P5, Section~\ref{sec:arch-principles}), A2 scoped attenuated delegation (G2), A3 key rotation with continuity (G1), A4 peer trust attestation with formal aggregation (G3), A5 identity-bound payment credential (G5), A6 relay-only infrastructure (G4 and deployment footprint), and A7 the STRIDE residual profile as a summary security posture. Evidence superscripts separate measured marks (NostrAgent) from literature-sourced and protocol-analytic ones. Section~\ref{sec:eval-threats} discusses this asymmetry. A6 is a property with a cost: relay-only operation removes registration authorities and consensus infrastructure but gives up the global consistency anchored systems get in return (TP-1 and FM-15, both treated in Section~\ref{sec:evaluation}). MCP~\cite{anthropic2024mcp} and A2A~\cite{a2a2025spec} are excluded as compositional rather than competing: they standardize interaction, not identity (Section~\ref{sec:bg-ecosystem}).

Vaidhyanathan and Taibi's review of ten general-purpose agentic frameworks finds a maturing but fragmented landscape with limited memory models, basic reasoning, and uneven ecosystem support, and notes that security remains largely underdeveloped~\cite{vaidhyanathan2026agentic}. NostrAgent addresses identity-layer deficits outside that framework-centric scope (G1 through G3), while G4 and G5 address the deployment seams between organizations. Framework-internal gaps such as memory and reasoning are orthogonal to this substrate layer. NostrAgent instantiates the Self-Sovereign Identity layer of the Sovereign-by-Design reference architecture~\cite{esposito2026sovereign} for agent identity.

\subsection{Centralized Approaches}\label{sec:rw-centralized}

OAuth~2.1~\cite{ietf2026oauth21} remains the standard mechanism for agent authorization. Tokens are issued by an identity provider (IdP), binding agent identity to a centralized authority and violating sovereignty by design: discovery routes through the provider (G4), and key continuity belongs to the provider rather than the operator (G1). The Agent Authorization Profile (AAP)~\cite{ietf2026aap} refines OAuth for multi-agent delegation with hop-counting \texttt{delegation} claims, but inherits the same IdP dependency, and its JSON Web Token (JWT) credentials do not support offline attenuation (G2 partial). The Agent Identity Protocol (AIP)~\cite{prakash2026aip} offers two modes: \texttt{aip:web:} anchors identity to DNS, while \texttt{aip:key:} uses self-certifying Ed25519 keys, architecturally equivalent to BIP340 keys. AIP's Biscuit/Datalog~\cite{couprie_biscuit} policy engine is more expressive than tag-based scoping (TP-5, Section~\ref{sec:eval-rq1}), and it ships a 600-test adversarial suite and an IETF draft~\cite{prakash2026aip,aip_ietf2026}. Even so, \texttt{aip:key:} provides no key rotation (G1), relay-based discovery (G4), peer attestation (G3), or payment binding (G5). SAGA~\cite{syros2026saga} provides the strongest centralized governance model, with ProVerif-verified secrecy and authentication properties (A7), but requires a centralized Provider as sole trust root.

\subsection{DID-Family Approaches}\label{sec:rw-did}

The W3C DID specification~\cite{w3c2022did} provides a general-purpose decentralized identifier framework, but key management varies by method: \texttt{did:web} depends on DNS, while \texttt{did:key} is self-certifying yet offers no rotation (G1). Composing delegation and attestation requires layering separate specifications (Verifiable Credentials~\cite{w3c2022vc} for attestation, ZCAP-LD~\cite{w3cccg2022zcapld} for authorization), with no unified delegation chain supporting scope attenuation (G2). Recent DID+VC proposals for agents~\cite{dids_vcs_agents2026} inherit the rotation gap. UCAN~\cite{ucan_spec_2025} provides attenuated capability delegation as DID-signed bearer envelopes (G2) but lacks identity-layer key rotation (G1), peer attestation (G3), and payment binding (G5). KERI~\cite{smith2019keri} provides the strongest key rotation in the comparison through witness consensus, pre-rotation, and duplicity detection via the Key Event Log (G1), and its companion specification ACDC~\cite{toip2023acdc} supports chained delegation with scope attenuation (G2). NostrAgent's pre-rotation is deliberately simpler than KERI's, omitting witness networks and trading duplicity-detection strength (FM-15) for relay-only infrastructure, agent-specific event semantics, and payment integration.

\subsection{Blockchain and Nostr-Native}\label{sec:rw-blockchain-nostr}

Blockchain-anchored approaches put identity and reputation on-chain. ERC-8004~\cite{erc8004_2025} deploys agent identity and reputation registries across EVM chains at per-operation gas cost, with no scoped delegation chains (G2). Layering ERC-8183~\cite{erc8183_2026} agentic-commerce primitives and x402~\cite{reppel2025x402} payments atop such registries approaches G5 but remains multi-layer. Within the Nostr ecosystem, Clawstr~\cite{clawstr2026} and the agentstr SDK~\cite{agentstr_sdk}, both layered over standard event kinds, specify neither scoped delegation chains (G2), peer attestation (G3), nor key rotation (G1). BitSov~\cite{larsen2026bitsov} defines a Bitcoin-native sovereignty stack at the transport and service layer. NostrAgent is complementary at the agent identity and delegation layer, using the same BIP340 and Lightning primitives. Three contemporary proposals extend this landscape: Vaziry et al.~\cite{vaziry2025agenteconomies} pair A2A AgentCards with x402 micropayments anchored on a ledger; Saavedra~\cite{saavedra2026delegation} formalizes delegation grants with optional blockchain anchoring; and Huang et al.~\cite{huang2025zerotrust} combine DIDs, VCs, and ZKP attestation in a zero-trust design. None combines relay-native identity, attenuated delegation, peer attestation, and Lightning-settled payment on one substrate.

Table~\ref{tab:comparison} summarizes: NostrAgent is the only row with full marks on A1 through A6, KERI/ACDC next closest, SAGA leading on A7 via ProVerif.

\begin{table}[!htbp]
\caption{Comparison of agent identity approaches.
  \checkmark~=~Full, $\circ$~=~Partial, $\times$~=~None.
  Evidence: $^M$~measured, $^L$~literature, $^P$~protocol analysis, $^A$~architectural argument.}
\label{tab:comparison}
\centering
\footnotesize
\setlength{\tabcolsep}{3.2pt}
\renewcommand{\arraystretch}{1.00}
\begin{tabular}{@{}lccccccc@{}}
\toprule
\textbf{System} &
\textbf{A1} &
\textbf{A2} &
\textbf{A3} &
\textbf{A4} &
\textbf{A5} &
\textbf{A6} &
\textbf{A7} \\
\midrule
\textbf{NostrAgent} &
  \checkmark$^{M\!,P}$ &
  \checkmark$^{M\!,P}$ &
  \checkmark$^{M\!,P}$ &
  \checkmark$^{M\!,P}$ &
  \checkmark$^{M\!,P}$ &
  \checkmark$^{A\!,P}$ &
  $\circ^{M\!,P}$ \\[2pt]
AIP (\texttt{key:}) &
  \checkmark$^{P\!,L}$ &
  \checkmark$^{L}$ &
  $\times$$^{P}$ &
  $\times$$^{P}$ &
  $\times$$^{P}$ &
  \checkmark$^{P}$ &
  $\circ^{L}$ \\
AIP (\texttt{web:}) &
  $\times$$^{P}$ &
  \checkmark$^{L}$ &
  $\circ^{P}$ &
  $\times$$^{P}$ &
  $\times$$^{P}$ &
  $\times$$^{P}$ &
  $\circ^{L}$ \\
KERI/ACDC &
  \checkmark$^{P\!,L}$ &
  \checkmark$^{P\!,L}$ &
  \checkmark$^{P\!,L}$ &
  $\times$$^{P}$ &
  $\times$$^{P}$ &
  $\circ^{P\!,L}$ &
  $\circ^{A}$ \\
W3C DID (\texttt{did:web}) &
  $\circ^{P\!,L}$ &
  $\circ^{P}$ &
  $\circ^{P}$ &
  $\circ^{P}$ &
  $\times$$^{P}$ &
  $\circ^{P}$ &
  $\times$$^{A}$ \\
ERC-8004 &
  $\circ^{P\!,L}$ &
  $\times$$^{P}$ &
  $\circ^{P}$ &
  $\circ^{P\!,L}$ &
  $\circ^{P}$ &
  $\times$$^{P}$ &
  $\times$$^{A}$ \\
Clawstr/agentstr &
  \checkmark$^{P}$ &
  $\times$$^{P}$ &
  $\times$$^{P}$ &
  $\times$$^{P}$ &
  $\circ^{P}$ &
  \checkmark$^{P}$ &
  $\times$$^{A}$ \\
SAGA &
  $\times$$^{P\!,L}$ &
  $\circ^{L}$ &
  $\circ^{L}$ &
  $\circ^{L}$ &
  $\times$$^{P}$ &
  $\times$$^{P\!,L}$ &
  \checkmark$^{L}$ \\
AAP (IETF) &
  $\times$$^{P}$ &
  $\circ^{L}$ &
  $\circ^{P}$ &
  $\times$$^{P}$ &
  $\times$$^{P}$ &
  $\times$$^{P}$ &
  $\times$$^{L}$ \\
OAuth 2.1 &
  $\times$$^{P\!,L}$ &
  $\circ^{P}$ &
  $\circ^{P}$ &
  $\times$$^{P}$ &
  $\times$$^{P}$ &
  $\times$$^{P}$ &
  $\times$$^{L}$ \\
\bottomrule
\end{tabular}
\vspace{2pt}

\noindent\scriptsize
A1~Decentralized identity; A2~Scoped attenuated delegation; A3~Key rotation with continuity; A4~Peer trust attestation with formal aggregation; A5~Identity-bound payment credential; A6~Relay-only infrastructure; A7~STRIDE residual profile.
\end{table}

The marks in Table~\ref{tab:comparison} record design trade-offs, not a ranking.
Centralized systems (OAuth, AAP, SAGA) buy governance, and in SAGA's case machine-checked security, at the cost of an identity provider, which is exactly G4 and the operator-sovereignty predicates.
KERI/ACDC is strongest on key continuity and attenuated delegation, but requires witness infrastructure that the relay-only choice (A6) forgoes, leaving FM-15 as the residual.
AIP's self-certifying mode matches our key model and exceeds our policy language (TP-5), yet has no rotation, attestation, or payment binding.
Blockchain registries buy global consistency that relays cannot, at per-operation cost and without scoped chains.
Nostr-native agent work inherits the relay substrate but does not specify G1-G3.
NostrAgent's bet is the opposite of KERI's: simpler infrastructure, agent-specific event semantics, and identity-bound payment, in exchange for no duplicity detection and no global log.
Section~\ref{sec:evaluation} measures that cost.

\section{NostrAgent Architecture}\label{sec:architecture}

\subsection{Design Principles}\label{sec:arch-principles}

The architecture rests on five operator-sovereignty predicates that formalize the operator's control over agent identity.
\textbf{P1 (Existence Independence):} an agent's keypair can be created without interaction with any external service.
\textbf{P2 (Verification Independence):} any party verifies identity and delegation chains using only the public key and signed events, without contacting a third party.
\textbf{P3 (Infrastructure Portability):} identity, delegation chains, and attestations migrate to any compatible relay without re-registration.
\textbf{P4 (Operator Supremacy):} the operator can unilaterally revoke, rotate, or re-scope identity and delegations.
\textbf{P5 (Non-Forgery by Infrastructure):} no relay or Lightning node can forge or modify identity without the private key.
These predicates align with the sovereignty conceptualization of Esposito et al.~\cite{esposito2026sovereign}: Controllability relates to P4, Autonomy to P1 and P3, Verifiability to P2 and P5.

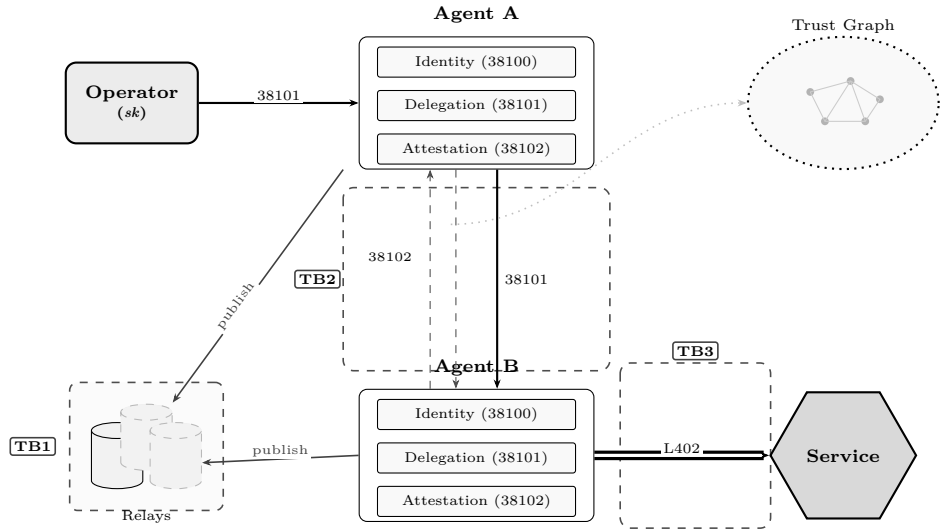
\begin{figure}[!htbp]
\centering
\begin{tikzpicture}[
  x=1cm, y=1cm,
  scale=0.97, transform shape,
  font=\fontsize{6.5}{8}\selectfont,
  >={Stealth[length=3.5pt]},
  operator/.style={
    draw, thick, rounded corners=4pt, fill=gray!15,
    minimum width=1.8cm, minimum height=1.1cm,
    align=center, font=\fontsize{7}{8.5}\selectfont\bfseries
  },
  agent/.style={
    draw, rounded corners=3pt, fill=white,
    minimum width=3.2cm, minimum height=1.8cm,
    align=center
  },
  slot/.style={
    draw, thin, fill=gray!5, rounded corners=1pt,
    minimum width=2.7cm, minimum height=0.32cm,
    font=\fontsize{5.5}{7}\selectfont, align=center
  },
  relay/.style={
    cylinder, draw, shape border rotate=90,
    minimum width=0.7cm, minimum height=0.9cm,
    fill=gray!10, font=\fontsize{5.5}{7}\selectfont,
    cylinder uses custom fill, cylinder end fill=gray!18,
    cylinder body fill=gray!8
  },
  service/.style={
    draw, thick, fill=gray!30,
    regular polygon, regular polygon sides=6,
    minimum size=1.5cm, align=center,
    font=\fontsize{7}{8.5}\selectfont\bfseries
  },
  trustcloud/.style={
    draw, dotted, thick, ellipse,
    minimum width=2.6cm, minimum height=1.8cm,
    fill=gray!5,
    font=\fontsize{6.5}{8}\selectfont, align=center
  },
  deleg/.style={->, thick, black},
  attest/.style={->, dashed, gray!60!black},
  l402arrow/.style={->, very thick, double, double distance=1.2pt, black},
  publish/.style={->, semithick, gray!50!black},
  tblabel/.style={
    font=\fontsize{5.5}{7}\selectfont\bfseries,
    fill=white, inner sep=1.5pt, draw=black!75, semithick,
    rounded corners=1pt
  },
  flowlabel/.style={
    font=\fontsize{5.5}{7}\selectfont,
    fill=white, inner sep=1pt, outer sep=0pt
  }
]

%


\node[operator] (op) at (0.8, 3.0)
  {Operator\\[-1pt]{\fontsize{5.5}{7}\selectfont(\textit{sk})}};

\node[agent] (agentA) at (5.5, 3.0) {};
\node[font=\fontsize{7}{8.5}\selectfont\bfseries, anchor=south]
  at ([yshift=2pt]agentA.north) {Agent A};
\node[slot] (a-id)    at ([yshift=-0.35cm]agentA.north) {Identity (38100)};
\node[slot] (a-deleg) at ([yshift=-0.38cm]a-id.south)   {Delegation (38101)};
\node[slot] (a-att)   at ([yshift=-0.38cm]a-deleg.south) {Attestation (38102)};

\node[trustcloud] (tg) at (10.5, 3.0) {};
\node[font=\fontsize{6}{7}\selectfont, anchor=south] at ([yshift=-2pt]tg.north)
  {Trust Graph};
\coordinate (gn1) at ([xshift=-0.45cm, yshift=0.15cm]tg.center);
\coordinate (gn2) at ([xshift=0.10cm, yshift=0.30cm]tg.center);
\coordinate (gn3) at ([xshift=0.50cm, yshift=0.05cm]tg.center);
\coordinate (gn4) at ([xshift=-0.25cm, yshift=-0.25cm]tg.center);
\coordinate (gn5) at ([xshift=0.30cm, yshift=-0.25cm]tg.center);
\foreach \n in {gn1,gn2,gn3,gn4,gn5}{
  \fill[gray!60] (\n) circle (1.5pt);
}
\draw[gray!40, thin] (gn1) -- (gn2) -- (gn3) -- (gn5) -- (gn4) -- (gn1);
\draw[gray!40, thin] (gn2) -- (gn5);
\draw[gray!40, thin] (gn4) -- (gn2);


\node[relay] (r1) at (0.6, -1.9) {};
\node[relay, draw=gray!55, dashed] (r2) at (1.0, -1.65) {};
\node[relay, draw=gray!55, dashed] (r3) at (1.4, -1.9) {};
\node[font=\fontsize{5.5}{7}\selectfont, anchor=north]
  at ([yshift=-5pt]r1.south -| r2) {Relays};

\node[agent] (agentB) at (5.5, -1.8) {};
\node[font=\fontsize{7}{8.5}\selectfont\bfseries, anchor=south]
  at ([yshift=2pt]agentB.north) {Agent B};
\node[slot] (b-id)    at ([yshift=-0.35cm]agentB.north) {Identity (38100)};
\node[slot] (b-deleg) at ([yshift=-0.38cm]b-id.south)   {Delegation (38101)};
\node[slot] (b-att)   at ([yshift=-0.38cm]b-deleg.south) {Attestation (38102)};

\node[service] (svc) at (10.5, -1.8) {Service};


\draw[deleg]
  (op.east) -- node[flowlabel, above, pos=0.5] {38101} (agentA.west);

\draw[publish]
  ([xshift=-6pt]agentA.south west) --
  node[flowlabel, above left=-1pt, sloped, pos=0.50,
    font=\fontsize{5}{6}\selectfont] {publish}
  ([yshift=6pt]r2.north east);

\draw[deleg]
  ([xshift=8pt]agentA.south) --
  node[flowlabel, right=2pt, pos=0.5] {38101}
  ([xshift=8pt]agentB.north);

\draw[attest]
  ([xshift=-8pt]agentA.south) -- ([xshift=-8pt]agentB.north);
\draw[attest]
  ([xshift=-18pt]agentB.north) -- ([xshift=-18pt]agentA.south);
\node[flowlabel, font=\fontsize{5}{6}\selectfont, anchor=east]
  at ([xshift=-24pt]$(agentA.south)!0.40!(agentB.north)$) {38102};

\draw[dotted, gray!50, semithick, ->]
  ([xshift=-10pt]$(agentA.south)!0.25!(agentB.north)$) to[out=0, in=180]
  (tg.west);

\draw[l402arrow]
  (agentB.east) -- node[flowlabel, above, pos=0.5] {L402} (svc.west);

\draw[publish]
  (agentB.west) --
  node[flowlabel, above, pos=0.50, font=\fontsize{5}{6}\selectfont] {publish}
  ([yshift=0pt]r3.east);


\begin{scope}[on background layer]
  \node[draw=black!70, dashed, semithick, rounded corners=3pt, inner sep=8pt,
        fit=(r1)(r2)(r3), fill=gray!3] (tb1fit) {};

  \draw[black!70, dashed, semithick, rounded corners=4pt]
    ([xshift=-6pt, yshift=-7pt]agentA.south west)
    rectangle
    ([xshift=6pt, yshift=7pt]agentB.north east);

  \draw[black!70, dashed, semithick, rounded corners=4pt]
    ([xshift=10pt, yshift=10pt]agentB.north east)
    rectangle
    ([xshift=-10pt, yshift=-10pt]svc.south west);
\end{scope}

\node[tblabel] at ([xshift=-14pt, yshift=0pt]tb1fit.west) {TB1};
\node[tblabel] at
  ([xshift=-16pt]$(agentA.south west)!0.50!(agentB.north west)$) {TB2};
\node[tblabel] at
  ($(agentB.north east)!0.50!(svc.north west) + (0, 0.65)$) {TB3};

\end{tikzpicture}
\caption{NostrAgent component-and-connector view with trust boundaries TB1 to TB3. Solid arrows: delegation (Kind~38101); dashed: attestation (Kind~38102); double: L402 payment. Relay borders: solid = operator-controlled, dashed = public.}
\label{fig:architecture}
\end{figure}

Figure~\ref{fig:architecture} shows the component-and-connector view on the running example. The operator, holding the identity secret key, issues a root delegation (Kind~38101) to agent A. A then sub-delegates a narrowed scope to B, and the two agents exchange peer attestations (Kind~38102, dashed) whose accumulation forms the emergent trust graph. Both agents publish their events to a relay set mixing an operator-controlled relay (solid border) with public relays (dashed). Relays store and forward but cannot forge. B pays the service through L402 (double arrow). Three trust boundaries anchor the STRIDE analysis (Section~\ref{sec:eval-rq3}): TB1 agent-relay, TB2 agent-agent, TB3 agent-service.


\subsection{Agent Identity (Kind 38100)}\label{sec:arch-identity}

Kind 38100 is parameterized replaceable, so each agent identity has a single canonical declaration at any time. The operator-chosen \texttt{d}-tag serves as a stable agent identifier (e.g., \texttt{weather-agent-v1}), the \texttt{p}-tag references the operator's public key, and one or more \texttt{t}-tags declare relay-filterable capabilities such as \texttt{weather-forecast}. The JSON content carries the agent's name, version, capability list, MCP and A2A endpoint URLs, and trust policy parameters. A \texttt{next\_key\_hash} tag commits the agent to a specific successor key for pre-rotation (Section~\ref{sec:arch-rotation}). Agents and verifiers discover identities through NIP-01 relay filters selecting by event kind, author public key, \texttt{d}-tag, or capability \texttt{t}-tag. In the running example, B finds the data service S by filtering for its capability tag. Operators choose their own relay set, and multi-relay publishing makes discovery a query that no single relay can forge or suppress.

\subsection{Delegation Chains (Kind 38101)}\label{sec:arch-delegation}

A Kind 38101 event encodes a scoped delegation from a delegator to a delegatee. Each event specifies a scope (capabilities, resources, actions), constraints (expiration, maximum chain depth, current depth), and a reference to the parent delegation or identity event via an \texttt{a}-tag.

Verification walks the chain from leaf to root, checking seven invariants at each hop: \textbf{INV-1 (Capability)} child capabilities $\subseteq$ parent capabilities; \textbf{INV-2 (Resource)} child resources $\subseteq$ parent resources (empty set denotes unrestricted); \textbf{INV-3 (Action)} child actions $\subseteq$ parent actions (empty set denotes unrestricted); \textbf{INV-4 (Depth)} current depth $\leq$ maximum chain depth; \textbf{INV-5 (Time)} child expiration $\leq$ parent expiration; \textbf{INV-6 (Continuity)} delegatee of hop $n$ equals author of hop $n{+}1$; \textbf{INV-7 (Acyclicity)} no cycles, enforced by a visited-set check. Together these guarantee per-hop narrowing at every link. For capabilities, which must be non-empty at every hop, narrowing is additionally transitive end to end. Attenuation is verifier-enforced rather than cryptographically bound at the signer, trading macaroon-style HMAC chaining for the simpler relay-replaceable model.

The empty-set convention in INV-2 and INV-3 creates an asymmetry. It exists for root-delegation ergonomics: an operator granting initial authority typically constrains capabilities but not yet every resource those capabilities may touch. Exhaustive root resource lists would invite wildcards with the same effect and less clarity. The consequence: because subset checks are per hop, a chain passing through an unrestricted intermediate hop does not preserve earlier resource or action restrictions. Capability narrowing, by contrast, is transitive because empty capability sets are rejected at creation. Section~\ref{sec:eval-discussion} discusses when this matters.

\begin{figure}[!htbp]
\centering
\begin{tikzpicture}[
    font=\fontsize{7}{8.4}\selectfont,
    msg/.style={-{Stealth[length=2.5pt]}, thick},
    selfmsg/.style={-{Stealth[length=2.5pt]}, thick},
    lifelinestyle/.style={dashed, gray!50},
    every node/.style={inner sep=1.5pt},
    x=1cm, y=1cm,
    scale=0.94, transform shape
]

\def\xOp{0}
\def\xA{2.4}
\def\xB{4.8}
\def\xV{7.2}

\node[draw, rounded corners=2pt, minimum width=1.3cm, minimum height=0.45cm,
      fill=gray!15, font=\fontsize{7}{8.4}\selectfont\bfseries]
      (hOp) at (\xOp, 0) {Operator};
\node[draw, rounded corners=2pt, minimum width=1.3cm, minimum height=0.45cm,
      fill=white, font=\fontsize{7}{8.4}\selectfont\bfseries]
      (hA) at (\xA, 0) {Agent A};
\node[draw, rounded corners=2pt, minimum width=1.3cm, minimum height=0.45cm,
      fill=white, font=\fontsize{7}{8.4}\selectfont\bfseries]
      (hB) at (\xB, 0) {Agent B};
\node[draw, rounded corners=2pt, minimum width=1.3cm, minimum height=0.45cm,
      fill=gray!8, font=\fontsize{7}{8.4}\selectfont\bfseries]
      (hV) at (\xV, 0) {Verifier};

\draw[lifelinestyle] (\xOp, -0.225) -- (\xOp, -5.55);
\draw[lifelinestyle] (\xA, -0.225)  -- (\xA, -5.55);
\draw[lifelinestyle] (\xB, -0.225)  -- (\xB, -5.55);
\draw[lifelinestyle] (\xV, -0.225)  -- (\xV, -5.55);


\def\yMone{-0.65}
\draw[msg] (\xOp, \yMone) -- node[above, font=\fontsize{6}{7.2}\selectfont]
    {delegate(\{a,b,c\}, depth=1/3)} (\xA, \yMone);

\def\yMthree{-1.35}
\draw[msg] (\xA, \yMthree) -- node[above, font=\fontsize{6}{7.2}\selectfont]
    {delegate(\{a,b\}, depth=2/3)} (\xB, \yMthree);

\def\yMfive{-2.05}
\draw[msg] (\xB, \yMfive) -- node[above, font=\fontsize{6}{7.2}\selectfont]
    {present chain} (\xV, \yMfive);

\draw[selfmsg] (\xV, -2.25) -- ++(0.6, 0) -- ++(0, -0.4) -- ++(-0.6, 0);
\node[right, font=\fontsize{6}{7.2}\selectfont, text=black]
    at (\xV + 0.08, -2.35) {verify\_chain()};

\def\yMseven{-2.8}
\draw[msg] (\xV, \yMseven) -- node[above, font=\fontsize{6}{7.2}\selectfont]
    {\textsc{ok}} (\xB, \yMseven);


\draw[decorate, decoration={brace, amplitude=3pt, mirror},
      thick, gray!70]
    (8.4, \yMone + 0.05) -- (8.4, \yMthree - 0.05);

\node[anchor=west, align=left, font=\fontsize{6}{7.2}\selectfont,
      inner sep=2pt]
    at (8.6, -0.65) {\{a,b,c\}\ \ depth: 1/3};

\node[anchor=west, font=\fontsize{6}{7.2}\selectfont, gray!70]
    at (8.6, -1.0) {$\downarrow\;\subseteq$ (attenuation)};

\node[anchor=west, align=left, font=\fontsize{6}{7.2}\selectfont,
      inner sep=2pt]
    at (8.6, -1.35) {\{a,b\}\ \ depth: 2/3};


\def\yDiv{-3.15}
\draw[dashed, gray!50, semithick]
    (\xOp - 0.8, \yDiv) -- (\xV + 0.8, \yDiv);
\node[fill=white, inner sep=1.5pt, font=\fontsize{6}{7.2}\selectfont\itshape,
      text=gray!70]
    at ({(\xA + \xB)/2}, \yDiv) {Revocation};

\def\yRone{-3.6}
\draw[msg] (\xOp, \yRone) -- node[above, font=\fontsize{6}{7.2}\selectfont]
    {revoke(d-tag, status=revoked)} (\xA, \yRone);
\node[font=\fontsize{6}{7.2}\selectfont\itshape, gray!60, anchor=west]
    at (\xA + 0.1, \yRone - 0.15) {(relay)};

\def\yRtwo{-4.1}
\draw[msg] (\xB, \yRtwo) -- node[above, font=\fontsize{6}{7.2}\selectfont]
    {present chain} (\xV, \yRtwo);

\draw[selfmsg] (\xV, -4.3) -- ++(0.6, 0) -- ++(0, -0.4) -- ++(-0.6, 0);
\node[right, font=\fontsize{6}{7.2}\selectfont, text=black]
    at (\xV + 0.08, -4.4) {ancestor revoked};

\def\yRfour{-4.9}
\draw[msg] (\xV, \yRfour) -- node[above, font=\fontsize{6}{7.2}\selectfont]
    {\textsc{reject}} (\xB, \yRfour);

\end{tikzpicture}
\caption{Delegation chain with scope attenuation (top) and cascade revocation (bottom).}
\label{fig:delegation-sequence}
\end{figure}
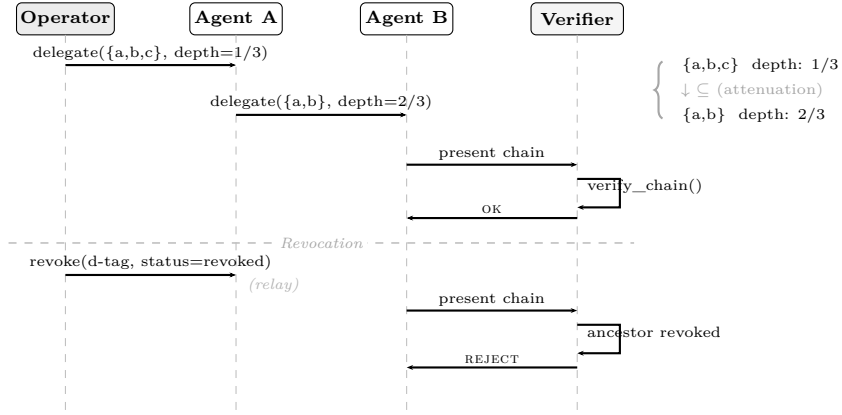

Like Kind 38100, Kind 38101 is parameterized replaceable, so cascade revocation works in-band: publishing a replacement with \texttt{revocation\_status} set to \texttt{"revoked"} invalidates that event and all downstream delegations, since any subsequent chain walk encounters the revoked ancestor and halts. Unlike bearer macaroons~\cite{birgisson2014macaroons} with external revocation state, Kind~38101 delegations revoke in-band and are non-transferable: INV-6 binds each hop to a specific delegatee public key, so the credential cannot be reused by any other party. Figure~\ref{fig:delegation-sequence} traces both halves: the operator delegates $\{a,b,c\}$ at depth 1 of 3 to A; A passes the narrowed $\{a,b\}$ to B; the verifier accepts after walking the invariants. Below, the operator revokes the root by republishing its \texttt{d}-tag with revoked status, and the next walk rejects.

\subsection{Peer Attestation and Trust (Kind 38102)}\label{sec:arch-trust}

To let agents form a transitive trust graph independent of relay or operator authority, Kind 38102 records a signed peer attestation: agent $A$ attests that agent $B$ possesses capability $C$ with confidence $w \in [0,1]$, serving as the edge weight in a directed trust graph. Trust computation uses depth-first search from the querying agent's operator-configured trust roots (anchors). Each path between $A$ and $B$ contributes independent evidence, combined via a noisy-OR gate~\cite{pearl1988probabilistic}: $\mathrm{trust}(A,B) = 1 - \prod_{p \in \mathcal{P}_{A \to B}} (1 - d^{|p|}\prod_{i} w_i)$, with $|p|$ the path edge count, defaults $d{=}0.5$ and depth $\leq 4$, and paths below $\epsilon{=}0.01$ pruned. The score is bounded in $[0,1]$ and monotone in path weights. The cost of the independence assumption is bounded analytically (Section~\ref{sec:eval-discussion}). L402's per-identity cost floor makes the scheme Sybil-deterrent through economic friction rather than formally Sybil-resistant~\cite{douceur2002sybil}. Attestations are strictly positive to prevent weaponized reputation attacks. Like Kinds 38100 and 38101, Kind 38102 is parameterized replaceable, keyed by the deterministic (attester, attestee, capability) d-tag, so attesters update confidence weights or revoke attestations in-band by republishing. This graph is how B chooses among competing weather services: attestations from A's anchors outweigh a newcomer's self-description.

\subsection{Key Rotation with Pre-Rotation}\label{sec:arch-rotation}

Pre-rotation~\cite{smith2019keri} commits to a successor key before it is used, defending against compromise of the active key. At identity creation, \texttt{next\_key\_hash} commits \texttt{SHA256(next\_pubkey)} over the 32-byte x-only encoding. At rotation, the old key publishes status \texttt{rotated} and the new key publishes status \texttt{active} with a \texttt{prev\_key} tag and a \texttt{rotation\_proof}: a BIP340 signature by the new key over \texttt{SHA256(old\_pk $\|$ new\_pk $\|$ timestamp)}. Verifiers reject any rotation in which \texttt{SHA256(new\_pk)} differs from the prior \texttt{next\_key\_hash}. A grace period (default 3600\,s) for relay propagation precedes old-key event rejection. This lets O rotate a suspected-compromised key without losing B's delegation or A's attestations.

\subsection{L402 Unified Identity-Payment}\label{sec:arch-l402}

\begin{figure}[!htbp]
\centering
\begin{tikzpicture}[
    >=Stealth,
    scale=0.94, transform shape,
    font=\scriptsize,                          
    msg/.style={font=\fontsize{6}{7.2}\selectfont},
    msgsub/.style={font=\fontsize{5.5}{6.6}\selectfont, text=black!45},
    note/.style={font=\fontsize{5.5}{6.6}\selectfont},
    phase/.style={font=\fontsize{6}{7.2}\selectfont\bfseries},
    lifeline/.style={dashed, gray!55},
    request/.style={->, semithick},
    response/.style={->, dashed},
  ]

  \def\xA{0}          
  \def\xS{4.0}        
  \def\xL{7.8}        
  \def\xP{9.4}        

  \fill[gray!5]  (-0.85, -3.65) rectangle (\xP+0.05, -4.55);
  \fill[gray!10] (-0.85, -4.85) rectangle (\xP+0.05, -5.85);

  \node[draw, rounded corners=2pt, minimum width=1.5cm, minimum height=0.45cm,
        fill=white, font=\scriptsize\bfseries] (agentH) at (\xA, 0) {Agent};
  \node[font=\fontsize{5}{6}\selectfont, text=gray!50]
    at (\xA, -0.36) {\textit{L402Client + Identity}};

  \node[draw, rounded corners=2pt, minimum width=1.5cm, minimum height=0.45cm,
        fill=gray!8, font=\scriptsize\bfseries] (serviceH) at (\xS, 0) {Service};

  \node[diamond, draw, minimum width=0.45cm, minimum height=0.45cm,
        fill=gray!18, font=\scriptsize\bfseries, inner sep=1pt] (lndH) at (\xL, 0) {LND};

  \def\yBot{-5.95}
  \draw[lifeline] (\xA, -0.27) -- (\xA, \yBot);
  \draw[lifeline] (\xS, -0.27) -- (\xS, \yBot);
  \draw[lifeline] (\xL, -0.35) -- (\xL, \yBot);

  \draw[decorate, decoration={brace, amplitude=3pt, mirror}]
    (\xP+0.1, -0.60) -- (\xP+0.1, -1.80)
    node[midway, right=4pt, phase] {Challenge};

  \draw[decorate, decoration={brace, amplitude=3pt, mirror}]
    (\xP+0.1, -3.65) -- (\xP+0.1, -4.55)
    node[midway, right=4pt, phase] {Payment};

  \draw[decorate, decoration={brace, amplitude=3pt, mirror}]
    (\xP+0.1, -4.85) -- (\xP+0.1, -5.85)
    node[midway, right=4pt, phase] {Access};


  \draw[request] (\xA+0.08, -0.80) -- (\xS-0.08, -0.80)
    node[midway, above, msg] {\texttt{GET /resource}};
  \node[msgsub] at ({(\xA+\xS)/2}, -0.95)
    {\texttt{X-Nostr-Pubkey: \textlangle pk\textrangle}};

  \draw[response] (\xS-0.08, -1.40) -- (\xA+0.08, -1.40)
    node[midway, above, msg] {\texttt{HTTP 402}};
  \node[msgsub] at ({(\xA+\xS)/2}, -1.55)
    {\texttt{macaroon + invoice}};

  \node[draw, rounded corners=3pt, note,
        fill=white, inner sep=3pt, align=center]
    (callout) at ({(\xA+\xS)/2}, -2.65)
    {\textbf{Macaroon ID} (65\,bytes):\\[1pt]
     \texttt{\fontsize{5}{6}\selectfont%
       ver(1B)\,$\|$\,pay\_hash(32B)\,$\|$\,pubkey(32B)}\\[2pt]
     {\small$\rightarrow$}\;\textit{Non-transferable: bound to BIP340 key}};
  \draw[gray!40, thin, ->] (callout.north) -- ++(0, 0.45);


  \draw[request] (\xA+0.08, -3.85) -- (\xL-0.12, -3.85)
    node[midway, above, msg] {\texttt{SendPayment(bolt11)}};

  \draw[response] (\xL-0.12, -4.35) -- (\xA+0.08, -4.35)
    node[midway, above, msg] {\texttt{preimage}};


  \draw[request, line width=1.0pt] (\xA+0.08, -5.05) -- (\xS-0.08, -5.05)
    node[midway, above, msg] {\texttt{GET + L402 mac:preimage}};
  \node[msgsub] at ({(\xA+\xS)/2}, -5.20)
    {\texttt{X-Nostr-Sig: \textlangle sig\textrangle}};

  \draw[response] (\xS-0.08, -5.65) -- (\xA+0.08, -5.65)
    node[midway, above, msg] {\texttt{HTTP 200 OK}};

\end{tikzpicture}
\caption{L402 unified identity-payment flow. The 65-byte macaroon identifier
binds the credential to the agent's BIP340 public key, making it
non-transferable.}
\label{fig:l402-flow}
\end{figure}
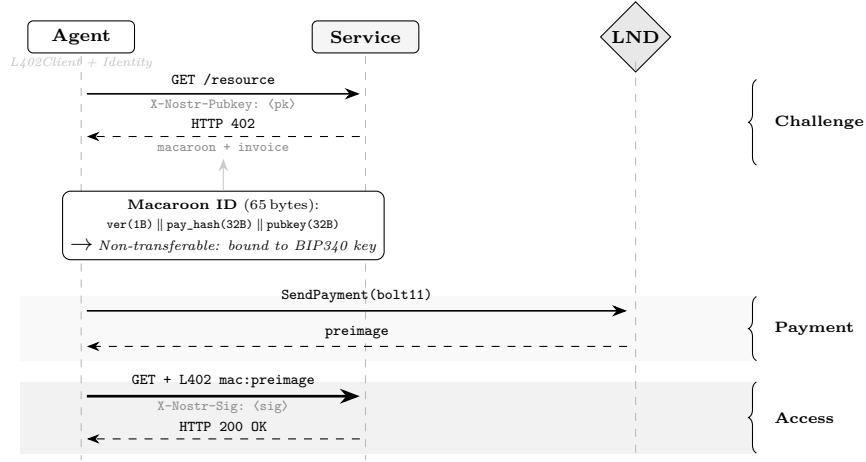

The L402 flow proceeds in three phases (Fig.~\ref{fig:l402-flow}). In the \emph{Challenge} phase, an agent presents its BIP340 public key to a service, which responds with HTTP~402 containing a macaroon and a Lightning invoice. The macaroon identifier is a 65-byte structure: version (1\,B), payment hash (32\,B), and agent public key (32\,B), binding the credential to the requesting agent. In the \emph{Payment} phase, the agent pays the invoice via its LND node and obtains the preimage. In the \emph{Access} phase, the agent resubmits the request with the macaroon, preimage, and a BIP340 signature over the canonical request (method, URL, timestamp, and body hash). The service performs five verification checks: preimage validity, identity match, BIP340 signature verification, timestamp freshness, and caveat satisfaction. The macaroon remains a bearer credential, but the preimage proves invoice settlement and the request signature proves key possession at access time. Together they authenticate the agent at verification rather than at issuance~\cite{birgisson2014macaroons}. Omitting the signature check degrades the credential to a bearer token, so deploying services must enforce it.

\subsection{Prototype}\label{sec:arch-prototype}

The prototype is implemented in Python~3.11 with nostr-sdk (Rust bindings) for BIP340, pymacaroons for L402, and LND on regtest. All three event kinds, key rotation with pre-rotation, and L402 payment verification are fully implemented. The test suite comprises 359 test functions spanning unit, property-based, adversarial STRIDE, and Docker integration layers. The benchmark testbed and environment are described in Section~\ref{sec:eval-setup}.

\subsection{What the Composition Buys}\label{sec:arch-emergent}

\textbf{Verification locality.} Every check, from BIP340 signatures through the seven invariants and trust aggregation to L402's five checks, is a pure computation over signed events plus the payment preimage. Retrieval is separable from checking: verifiers fetch events from any relay or local cache, then decide offline, contacting no issuing authority. S authorizes B's request without calling the operator, A, or any authorization server, in well under a millisecond of compute (B3, B11).

\textbf{Cross-layer key binding.} One BIP340 key signs the identity event, sits inside the 65-byte macaroon identifier, and signs each L402 request. Identity, authorization, and payment therefore hold or fail together: a stolen macaroon plus preimage is unreplayable because the third element, key possession, cannot be lifted from the event log (B11).

\textbf{Priced trust.} Free attestation edges and priced identities are individually weak defenses. Composed, they yield measurable Sybil deterrence (FM-8, FM-13). The property lives in neither module: the trust engine has no notion of payment, and L402 has no notion of trust (Section~\ref{sec:eval-rq2}).

\textbf{Lifecycle uniformity.} With all three kinds parameterized replaceable, rotation, revocation, and retraction are one republish-under-the-same-\texttt{d}-tag operation, so revocation cascades in-band with no revocation service (B8).

Read against the predicates and gaps of Sections~\ref{sec:arch-principles} and~\ref{sec:introduction}: P1, P2, and G4 are Kind~38100 plus NIP-01 filters; P4, G1, and G2 are pre-rotation plus the Kind~38101 invariants and in-band replaceability; G3 is Kind~38102 with the L402 cost floor; G5 is the 65-byte macaroon binding. P3 and P5 follow from relays being substitutable and unable to forge BIP340 events. The security and sovereignty properties claimed in Section~\ref{sec:introduction} are these mechanisms in combination, not any one of them.

\section{Evaluation and Discussion}\label{sec:evaluation}

This section answers RQ1 through RQ3 with a mixed-method design: ATAM scenario analysis with priorities rated by an expert panel (RQ1), benchmarks organized as a gap-fulfillment analysis (RQ2), and STRIDE enumeration plus adversarial failure-mode testing (RQ3). The methods are complementary: expert judgment establishes which qualities matter and how hard they are, benchmarks quantify the prioritized behaviors, and the security analyses probe the same architecture under attack rather than under load.

\subsection{Evaluation Setup}\label{sec:eval-setup}

All experiments run against the prototype (Section~\ref{sec:arch-prototype}) in one of three environments. Offline microbenchmarks measure pure verification computations in-process, with no network; each metric runs 1000 times, the first 100 discarded as warm-up ($N{=}900$). Docker benchmarks run against a live testbed of six Compose containers: three strfry Nostr relays, two LND Lightning nodes, and one bitcoind node in regtest mode (a private, deterministic Bitcoin network). These run 100 times each, with 10 warm-up discards for the stateful flows B7 to B9 ($N{=}90$). One failure-mode test (FM-11) publishes to five public Nostr relays over the open internet, the only measurement in this paper taken outside local infrastructure.

Table~\ref{tab:bench-defs} defines B1 to B11 and their environments; B6b is a parameter-sweep variant of B6. The trust-graph benchmarks build seeded Erd\H{o}s-R\'enyi~\cite{erdos1959random} synthetic graphs (edge probability 0.15).

Comparisons between NostrAgent configurations use two-sided Mann-Whitney U tests with Cliff's delta effect sizes~\cite{arcuri2011practical}. Every pairwise sweep comparison (chain depth B3, relay count B4, graph size B6, depth limit B6b, caveat count B10) is significant at $p < 0.001$ with a large effect ($|\delta| \geq 0.94$); per-pair results and exact dependency versions are in the replication package. All runs fix seed 42 (Apple Silicon macOS host, Python 3.11).

\begin{table}[t]
\centering
\caption{Benchmark definitions. Offline = in-process cryptographic microbenchmark; Docker = live Compose testbed. Run counts and warm-up protocol in Section~\ref{sec:eval-setup}; results in Table~\ref{tab:benchmarks}.}
\label{tab:bench-defs}
\footnotesize
\begin{tabular}{@{}lp{8.6cm}l@{}}
\toprule
ID & What it measures & Env. \\
\midrule
B1 & Kind~38100 verification: serialization, event-id recomputation, BIP340 verify, content parse & Offline \\
B2 & Isolated BIP340 signature verification, per event kind & Offline \\
B3 & Delegation-chain verification at depths 1 through 4 & Offline \\
B4 & Relay discovery of a Kind~38100 identity across 1 to 3 relays & Docker \\
B5 & Kind~38102 attestation validation: schema, event id, signature & Offline \\
B6 & Trust-graph query (noisy-OR depth-first search), 5 to 500 nodes & Offline \\
B6b & Trust-query variant of B6: depth-limit and decay sweeps & Offline \\
B7 & End-to-end key rotation published through the relays & Docker \\
B8 & Revocation propagation until all three relays serve the revoked event & Docker \\
B9 & Full L402 flow: challenge, Lightning payment, five-check verification & Docker \\
B10 & Macaroon attenuation cost at 1 to 20 caveats & Offline \\
B11 & Five-check L402 credential verification in isolation & Offline \\
\bottomrule
\end{tabular}
\end{table}

\begin{table}[!htbp]
\caption{NostrAgent benchmark results: representative medians per benchmark B1 to B11, with the full B3 depth and B6b $D_{\max}$ sweeps. B6 holds $D_{\max}{=}4$ (default); B6b sweeps $D_{\max}$ at fixed $n{=}50$. Environments per benchmark in Table~\ref{tab:bench-defs}.}
\label{tab:benchmarks}
\centering
\small
\setlength{\tabcolsep}{4pt}
\renewcommand{\arraystretch}{0.93}
\begin{tabular}{llrrrrc}
\toprule
ID & Variant & Median (ms) & IQR & P95 & P99 & N \\
\midrule
  B1  & kind\_38100 auth          & 0.0165  & 0.0005  & 0.0169  & 0.0204  & 900 \\
  B2  & BIP340 verify (all kinds) & 0.0154  & 0.0001  & 0.0157  & 0.0169  & 900 \\
  B3  & depth 1                   & 0.0155  & 0.0001  & 0.0159  & 0.0196  & 900 \\
  B3  & depth 2                   & 0.0308  & 0.0005  & 0.0320  & 0.0365  & 900 \\
  B3  & depth 3                   & 0.0460  & 0.0001  & 0.0487  & 0.0520  & 900 \\
  B3  & depth 4 ($\approx$15\,us/hop) & 0.0619  & 0.0002  & 0.0658  & 0.0698  & 900 \\
  B4  & discovery, 3 relays       & 2.3     & 0.20    & 2.7     & 2.8     & 100 \\
  B5  & kind\_38102 attestation   & 0.0168  & 0.0001  & 0.0174  & 0.0226  & 900 \\
  B6  & trust query, 5 nodes      & 0.0005  & 0.0000  & 0.0005  & 0.0015  & 900 \\
  B6  & trust query, 100 nodes    & 3.0     & 0.07    & 3.2     & 3.6     & 900 \\
  B6  & trust query, 500 nodes    & 2422.4  & 175.3   & 2918.3  & 3275.9  & 900 \\
  B6b & $D_{\max}{=}2$ ($n{=}50$) & 0.0100  & 0.0003  & 0.0104  & 0.0107  & 900 \\
  B6b & $D_{\max}{=}3$ ($n{=}50$) & 0.0659  & 0.0008  & 0.0688  & 0.0748  & 900 \\
  B6b & $D_{\max}{=}4$ ($n{=}50$) & 0.4487  & 0.0041  & 0.4602  & 0.4704  & 900 \\
  B7  & key rotation              & 235.8   & 7.9     & 277.6   & 289.8   & 90  \\
  B8  & revocation propagation    & 336.8   & 9.4     & 352.6   & 392.4   & 90  \\
  B9  & L402 full flow            & 157.4   & 9.6     & 166.7   & 170.1   & 90  \\
  B10 & macaroon, 10 caveats      & 0.0517  & 0.0005  & 0.0545  & 0.0598  & 900 \\
  B11 & L402 5-check verification & 0.0552  & 0.0004  & 0.0577  & 0.0633  & 900 \\
\bottomrule
\end{tabular}
\end{table}

\subsection{RQ1: Architectural Trade-offs}\label{sec:eval-rq1}

RQ1 asks how NostrAgent's design choices trade among sovereignty, operability, and trust. Following lightweight ATAM~\cite{kazman2000atam,sahlabadi2022lightweight}, informed by the six of AgentArcEval's eleven agent quality attributes~\cite{lu2026agentarceval} that bear on the identity layer, we selected twelve quality-attribute leaves and organized them into a three-branch utility tree (Table~\ref{tab:utility-tree}). Three independent anonymous experts in relevant fields rated each leaf on importance and difficulty in a two-round mini-Delphi~\cite{dalkey1963delphi}: a blind asynchronous rating round, then a synchronous convergence session. Round~2 agreement was near-perfect: Krippendorff's $\alpha = 0.94$~\cite{krippendorff2004content} averaged across axes (importance 1.00, difficulty 0.88; 95\% bootstrap CI [0.78, 1.00]), up from 0.79 in Round~1. Fleiss $\kappa = 0.78$~\cite{fleiss1971measuring} on joint (importance, difficulty) categories for ATAM comparability, up from 0.46 in Round~1. The panel reached unanimity on nine of the twelve leaves and left three persistent difficulty dissents (O2, T1, T4), flagged in Table~\ref{tab:utility-tree}.

\begin{table}[t]
\centering
\caption{ATAM utility tree with final Round~2 (importance, difficulty) ratings from the mini-Delphi panel; the bold \textbf{H,H} leaves are the differentiators anchoring RQ1.}
\label{tab:utility-tree}
\footnotesize
\begin{tabular}{@{}lllc@{}}
\toprule
Branch & Leaf & Concern & (Imp, Diff) \\
\midrule
Sovereignty & \textbf{S1} & Offline verifiability & \textbf{H,H} \\
            & S2 & Censorship resistance (relay migration) & H,M \\
            & \textbf{S3} & Key portability (rotation enforcement) & \textbf{H,H} \\
            & S4 & Infrastructure independence & M,L \\
\midrule
Operability & O1 & Verification latency & H,M \\
            & O2 & Framework interoperability$^{d}$ & M,M \\
            & O3 & Auditability (chain reconstruction) & H,M \\
            & O4 & Modifiability (schema extension) & M,L \\
\midrule
Trust       & T1 & Attestation validity (forgery rejection)$^{d}$ & H,L \\
            & \textbf{T2} & Trust graph integrity after rotation & \textbf{H,H} \\
            & \textbf{T3} & Sybil deterrence (ring bounding) & \textbf{H,H} \\
            & T4 & Trust bootstrapping (cold start)$^{d}$ & M,M \\
\bottomrule
\multicolumn{4}{@{}l@{}}{\scriptsize $^{d}$one persistent difficulty dissent (Sect.~\ref{sec:eval-rq1}).}
\end{tabular}
\end{table}

A sensitivity point (SP) is a parameter whose variation measurably shifts a leaf's response, each mapped to a backing benchmark or failure-mode test. A trade-off point (TP) is a decision in which improving one leaf demonstrably harms another. Four H,H leaves are the differentiators, each with a measured anchor: S1 offline verifiability (depth-4 chain verifies in 0.062~ms, B3), S3 key portability (rotation 236~ms, B7, old-key events rejected in all trials), T2 trust-graph integrity (attestations survive rotation), and T3 Sybil ring bounding (up to about 20{,}000 sats per fabricated trust unit, FM-13).

\textbf{TP-1 (relay federation: availability vs.\ consistency).} Multi-relay publishing buys censorship resistance (S2) and offline verifiability (S1) at the price of consistency: relays never synchronize, so verifiers can observe divergent event sets. B4 prices the added latency; FM-14 and FM-15 characterize the consistency cost.

\textbf{TP-2 (auditability vs.\ relationship privacy).} Kind~38102 publishes the attestation graph in cleartext: exactly what makes chains reconstructible (O3) and trust replayable, and what concedes relationship metadata to any observer. The cost holds by construction rather than by measurement, and Table~\ref{tab:stride} rates it the dominant Information Disclosure residual.

\textbf{TP-3 (single-level pre-rotation vs.\ propagation window).} Pre-rotation hardens S3 but opens a grace window (default 3600~s) while relays propagate. Measured via B7 and B8, probed by FM-6 and FM-12.

\textbf{TP-4 (L402 coupling).} Binding identity to Lightning settlement buys the G3 cost floor (FM-13) but couples the architecture to Lightning liveness. Measured via B9 (the round-trip cost), probed by FM-4 (clean recovery from an LND outage).

\textbf{TP-5 (tag-based scoping vs.\ expressiveness).} Tag scopes verify in microseconds offline with a small auditable verifier but cannot express temporal or relational constraints. AIP's Biscuit/Datalog engine is strictly more expressive. The empty-set convention that makes root delegations ergonomic also limits attenuation transitivity for resources and actions (Section~\ref{sec:arch-delegation}), a disclosed cost of simplicity. Measured via B10 and B11.

\textbf{Sensitivity points (SP-1 to SP-5, Table~\ref{tab:benchmarks}).} Each SP is a deployment knob, and the sweeps justify the defaults Section~\ref{sec:architecture} ships. Relay count moves resolution from 1.4 to 2.3~ms across one to three relays (SP-1, B4), so S2's relay migration is bounded by republishing (B7, B8) rather than discovery. Chain depth (O1) scales linearly at roughly 15 microseconds per hop (SP-2, B3). The depth limit $D_{\max}$ is the dominant cost knob, a 45$\times$ spread across $D \in \{2,3,4\}$ at $n{=}50$ (SP-3, B6b). The default 4 buys the widest attestation-path coverage for 0.45~ms, and because depth also extends Sybil-ring reach it is the first lever to lower under load. Pre-rotation (S3) rejects every competing rotation tested (SP-4, FM-6). Query cost plateaus at decay $d \geq 0.5$ (SP-5, B6b), where the $\epsilon{=}0.01$ prune stops truncating depth-4 paths, placing the default $d{=}0.5$ at the knee.

\subsection{RQ2: Does Unification Close the Gaps?}\label{sec:eval-rq2}

RQ2 asks whether unifying identity, trust, and payment on one substrate yields architectural advantages over composing separate systems. We answer against the five gaps of Section~\ref{sec:introduction}: Table~\ref{tab:gap-matrix} maps each gap to its mechanism and evidence, and the text walks the two gaps where unification does the most work.

\begin{table}[t]
\centering
\caption{Gap fulfillment: the five deficits of Section~\ref{sec:introduction}, the mechanism closing each, and the measured evidence.}
\label{tab:gap-matrix}
\footnotesize
\begin{tabular}{@{}p{2.2cm}p{4.1cm}p{5.2cm}@{}}
\toprule
Gap & Mechanism & Evidence \\
\midrule
G1 key continuity & Pre-rotation and \texttt{prev\_key} chain (Sect.~\ref{sec:arch-rotation}) & B7: rotation 236~ms; FM-6: competing rotations rejected; FM-7: continuity; attestations survive rotation (T2) \\
\addlinespace
G2 attenuated delegation & Kind~38101, INV-1 through INV-7 (Sect.~\ref{sec:arch-delegation}) & B3: linear verification at roughly 15~\textmu s per hop; B10: 4.2~\textmu s per caveat; B8: revocation on all relays in 337~ms; FM-9 and FM-17: escalation rejected \\
\addlinespace
G3 peer trust & Kind~38102 attestations, noisy-OR aggregation, L402 cost floor (Sect.~\ref{sec:arch-trust}) & FM-8: decay caps Sybil trust at 0.714 vs.\ 1.0; FM-13: 10{,}000 to 20{,}000 sats per unit of fabricated trust; B6: query scaling \\
\addlinespace
G4 registry-free discovery & Kind~38100, NIP-01 filters, multi-relay publishing (Sect.~\ref{sec:arch-identity}) & B4: resolution 1.4 to 2.3~ms (SP-1); FM-1: graceful degradation; FM-2: omission defeated \\
\addlinespace
G5 identity-bound payment & 65-byte macaroon identifier, preimage, BIP340 request signature (Sect.~\ref{sec:arch-l402}) & B9: settlement 157~ms; B11: five checks in 0.055~ms; FM-4: LND-outage recovery \\
\bottomrule
\end{tabular}
\end{table}

\textbf{G5: identity-bound payment.} The L402 flow of Section~\ref{sec:arch-l402} collapses identity verification and payment settlement into one HTTP exchange (Fig.~\ref{fig:l402-flow}). B9 reports 157~ms median end to end on regtest, dominated by the Lightning round-trip. B11 isolates the five-check verification at 0.055~ms. The preimage and identity checks bind only public values, so the BIP340 request signature is the sole proof of key possession.

\textbf{G3: priced peer trust.} Deterrence emerges from composing two individually weak mechanisms. Noisy-OR with decay caps what a ring of fake identities extracts: FM-8 bounds honest-to-Sybil trust at 0.714 versus 1.0 (28.6\% reduction at $d{=}0.5$, three bridges). Yet without a cost floor a 20-node ring still saturates its trust mass at 0.998 (FM-13). The 1000-sat-per-identity L402 floor prices that ring, raising fabrication cost from about 10{,}000 sats per trust unit at five identities to about 20{,}000 at saturation. The scheme is Sybil-deterrent through economic friction, not formally Sybil-resistant: identity cost is linear with no superlinear penalty, and clearing a decision threshold is far cheaper than saturating trust, since five identities already reach 0.48 for 5{,}000 sats.

G1, G2, and G4 close on the same substrate with the evidence listed in Table~\ref{tab:gap-matrix}. The notable single number is B8, where an in-band revocation becomes visible on all three relays in 337~ms median. Trust-query cost (B6, Table~\ref{tab:benchmarks}) grows steeply with graph size. The controlling knob is $D_{\max}$ (SP-3), and Section~\ref{sec:eval-discussion} turns the measured curve into deployment guidance. The answer to RQ2 is architectural: one key and one event log yield properties a composed stack cannot express, identity-bound payment (G5) and in-band cascade revocation (G2), with the head-to-head latency comparison left unmeasured (Section~\ref{sec:eval-threats}).

\subsection{RQ3: Failure Modes and Threat Analysis}\label{sec:eval-rq3}

RQ3 asks what failure modes arise in an operator-sovereign, relay-based architecture. The catalog enumerates 19 modes spanning relay availability, key management, trust-graph manipulation, Lightning unavailability, cross-relay consistency, and cryptographic correctness. 17 are empirical (11 offline cryptographic, 5 on the Docker testbed, and FM-11 against public relays) and all pass. FM-3 is an analytical redundancy bound, and FM-15 is an acknowledged architectural limitation (Section~\ref{sec:eval-discussion}). Table~\ref{tab:failure-modes} lists per-mode results. FM-3 bounds unrecoverable data loss at $(1-p)^R$ for $R$ independent relays ($\leq 10^{-6}$ at $R{=}3$, $p{=}0.99$), with FM-1 and FM-14 as empirical anchors.

\begin{table}[!htbp]
\caption{Failure mode results (FM-1 to FM-19). FM-3, FM-15 are analytical; the other 17 are empirical and pass. Related modes share a row.}
\label{tab:failure-modes}
\centering
\small
\setlength{\tabcolsep}{4pt}
\renewcommand{\arraystretch}{0.93}
\begin{tabular}{llp{7.4cm}}
\toprule
ID & Status & Detail \\
\midrule
  FM-1      & PASS       & Relay partition: resolution succeeds with 1/2/3 relays (3.9/3.3/11.1\,ms) \\
  FM-2      & PASS       & Relay malice: union queries defeat censorship-by-omission; BIP340 signatures preclude tampering \\
  FM-4      & PASS       & L402 fails cleanly on LND outage; recovers without channel corruption \\
  FM-5      & PASS       & Revocation verification: 33.4\,\textmu{}s \\
  FM-6      & PASS       & Key compromise: attacker rejected, legitimate rotation accepted \\
  FM-7      & PASS       & Key loss: identity chain valid after pre-rotation recovery \\
  FM-8      & PASS       & Trust poisoning: Sybil trust 0.714 vs honest 1.0 (28.6\% reduction, $d{=}0.5$, 3 bridges) \\
  FM-9, 17  & PASS       & Chain verification linear to depth 16 (FM-9); escalation rejected, valid narrowing accepted (FM-17) \\
  FM-10     & PASS       & Temporal validity: $\pm$60\,s tolerance enforced; 30/60\,s skew accepted, 120/300\,s skew rejected \\
  FM-11, 16 & PASS       & 3 of 5 public relays accepted all three kinds; the other 2 were unreachable, with no kind-based rejection observed (FM-11); 200-event flood publishes at 0.95\,ms median (FM-16) \\
  FM-12     & PASS       & Rotation: old-key orphan delegations invalidated; new-key re-issuance verifies \\
  FM-13     & PASS       & Sybil cost: linear per-identity, sub-linear marginal cost-per-trust-unit (5/10/20 identities) \\
  FM-14     & PASS       & Cross-relay consistency: no gossip; agents must multi-relay publish \\
  FM-3, 15  & ANALYTICAL & Multi-relay redundancy argument (FM-3); duplicity gap (FM-15) \\
  FM-18, 19 & PASS       & Offline sanity: 1000 unique keypairs with 0 collisions (FM-18); BIP340 signature validity (FM-19) \\
\bottomrule
\end{tabular}
\end{table}

Table~\ref{tab:stride} reports STRIDE residual risk across TB1 (Agent-Relay), TB2 (Agent-Agent), and TB3 (Agent-Service). BIP340 holds Spoofing and Tampering at 1 or 2 everywhere. Repudiation at TB1 stays a 3 because relays can deny receipt without counter-proof. The dominant residual is Information Disclosure (4 of 5 on TB1 and TB2): observers can reconstruct the attestation graph, intrinsic to the relay-only model (TP-2) rather than a mitigation gap.

\begin{table}[!htbp]
\centering
\caption{STRIDE residual-risk across TB1 (Agent-Relay), TB2 (Agent-Agent), TB3 (Agent-Service); 1 (negligible) to 5 (critical) post-mitigation.}
\label{tab:stride}
\small
\setlength{\tabcolsep}{6pt}
\begin{tabular}{@{}lccc@{}}
\toprule
\textbf{Category} & \textbf{TB1} & \textbf{TB2} & \textbf{TB3} \\
\midrule
Spoofing                & 2 & 2 & 2 \\
Tampering               & 2 & 1 & 2 \\
Repudiation             & 3 & 2 & 2 \\
Information Disclosure  & \textbf{4} & \textbf{4} & 3 \\
Denial of Service       & 3 & 2 & 3 \\
Elevation of Privilege  & 2 & 2 & 2 \\
\bottomrule
\end{tabular}
\end{table}

\subsection{Discussion}\label{sec:eval-discussion}

\textbf{Architectural limitation (FM-15).} NostrAgent assumes an honest majority in the operator-controlled relay set. Byzantine equivocation across disjoint relay subsets is undetectable without witness consensus (cf.\ KERI~\cite{smith2019keri}), which the relay-only principle forbids. Relay selection, superset verifier queries with NIP-01 tiebreaking, and pre-rotation bound the blast radius. Relay-only operation is likewise a poor fit where mandatory global consistency or regulated settlement applies.

\textbf{Operator key compromise.} Pre-rotation makes a stolen key recoverable rather than terminal: the attacker holds the active key but not the committed successor, so only the operator can rotate the identity (FM-6). Rotation orphans the delegations issued under the old key, which must be re-issued under the new one (FM-12). The compromised key stops verifying once the grace window closes, but inside that window its signatures still verify and conflicting publications to disjoint relay sets stay undetectable (FM-15).

\textbf{Attenuation transitivity.} An unrestricted intermediate hop does not preserve earlier resource or action restrictions, so end-to-end narrowing holds only for capabilities (Section~\ref{sec:arch-delegation}). The property-based suite pins the boundary, and FM-17 confirms that escalation against restricted parents is rejected. Deployments needing end-to-end resource narrowing should restrict every hop. Richer policy engines (TP-5) remove the sharp edge.

\textbf{Production-scale guidance.} At fixed edge probability the local scaling exponent rises with graph size, from about 1.7 (5 to 10 nodes) to about 4.5 (200 to 500), because expected degree grows with $n$. Interpolating the measured segments at $D_{\max}{=}4$: under 10~ms to roughly 140 nodes, under 100~ms to roughly 250, one second near 410. The depth lever comes first (dropping $D_{\max}$ from 4 to 3 cut latency 6.8$\times$ at $n{=}50$, B6b), then caching and precomputation. Sparser graphs shift the crossings outward, and the guidance is prototype-derived, not production-validated.

\textbf{Regulatory alignment.} Traceable delegation chains and attestations provide transparency artefacts aligned with the OWASP Agentic Top~10~\cite{owasp_agentic_top10_2026}, EU AI Act high-risk requirements~\cite{eu_ai_act_2024}, and the emerging profiles of the NIST National Cybersecurity Center of Excellence~\cite{nist_nccoe_agents2026} and IETF WIMSE~\cite{ietf_wimse_agent2026}.

\textbf{Concessions and assumptions.} TP-2 concedes relationship-metadata privacy for auditability (future work: pseudonymous attester keys, attestation-only relays); TP-5 concedes policy expressiveness for sub-millisecond offline verification. We assume secure operator key custody, one honest reachable relay, a classical adversary, and correct nostr-sdk and libsecp256k1 implementations. Side-channel and supply-chain attacks are out of scope. Single-level pre-rotation leaves simultaneous compromise of active and successor keys irrecoverable. Noisy-OR treats trust paths as independent, so when paths share a sub-path the shared support is counted once per path, an always-upward bias that is analytically bounded (any two-hop pair sharing one hop is overestimated by at most $1/16$). The empirical bias distribution is unmeasured and left to future work.

\subsection{Threats to Validity}\label{sec:eval-threats}

\textbf{Internal.} STRIDE residuals and comparison marks (Tables~\ref{tab:stride}, \ref{tab:comparison}) are author-assigned and carry researcher-bias risk.

\textbf{External.} The testbed uses 5 agents, 3 relays, regtest Lightning, and graphs capped at 500 nodes. Production-scale clustering and mainnet fee dynamics are unobserved. Regtest excludes real network propagation and routing fees.

\textbf{Construct.} The utility tree, STRIDE categories, and axes A1-A7 are author-curated. AgentArcEval and the standard STRIDE taxonomy ground the candidate pool, but leaf selection is author judgment; the experts rated the tree as given, so their agreement covers the ratings, not the tree's construction. NostrAgent marks are measured while baseline marks are literature-sourced or protocol-analytic, an asymmetry Table~\ref{tab:comparison} makes explicit through evidence superscripts. Likewise, RQ2's advantage over a composed multi-system baseline is argued analytically and positioned through that comparison. No composed baseline was measured head to head.

\section{Conclusion}\label{sec:conclusion}

We have shown that operator-sovereign agent identity, scoped delegation, peer-attested trust, registry-free discovery, and Lightning-gated payment can be unified on a relay-only substrate without a registration authority. NostrAgent closes the five gaps named in Section~\ref{sec:introduction}: key rotation preserves identity, delegations, and trust (G1, 236~ms median); every hop verifiably narrows the granted capabilities and revocation cascades in one event (G2); peer attestation over a priced identity floor deters Sybil fabrication (G3, up to about 20{,}000 sats per fabricated trust unit); discovery needs no registry (G4, 1.4 to 2.3~ms); and payment credentials bind to the paying agent, so a stolen macaroon is unreplayable (G5, settlement at 157~ms on regtest). The contribution is the composition: each primitive is mature in isolation, but one key type, one event log, and one revocation primitive yield offline verification of the whole authorization tuple, in-band cascade revocation, and identity-bound payment in one Lightning round-trip. The result runs on commodity relays and is safe in the tested sense of the catalog: 17 of 19 failure modes pass adversarial tests, FM-3 is bounded analytically, and FM-15 remains the honest limit of the relay-only model. The mixed-method evaluation substantiates C1 through C3, and the replication package delivers C4. Trust-query latency beyond 500 nodes and single-level pre-rotation's Byzantine-equivocation gap remain the visible limitations. Future work: mainnet L402 validation, large-scale trust simulation with empirical noisy-OR bias measurement and a sweep of the $\epsilon$ pruning threshold, formal verification of the attenuation invariants, and NIP standardization of Kinds 38100 through 38102.

\section*{Data Availability}

The replication package (implementation, benchmark harness, raw results, Docker testbed, tests, ATAM/mini-Delphi materials, failure-mode results) is archived at \url{https://doi.org/10.5281/zenodo.22744262} and mirrored at \url{https://github.com/Oliver1703dk/agentics2026-replication-package}. Experiments use fixed seed \texttt{42} and pinned dependencies.

\bibliographystyle{splncs04}
\bibliography{references}

\end{document}